\documentclass[11pt]{article}

\usepackage[preprint]{acl}

\usepackage{times}
\usepackage{latexsym}
\usepackage{booktabs} 
\usepackage{algorithm}
\usepackage{algorithmic}
\usepackage{booktabs}
\usepackage{float}
\usepackage{amsmath}
\usepackage{amssymb}
\usepackage[T1]{fontenc}

\usepackage[utf8]{inputenc}

\usepackage{microtype}

\usepackage{inconsolata}

\usepackage{graphicx}

\title{Who Responds When the Driver Is Gone? \\ A Framework for Holistic Passenger Intent Understanding}

\author{\textbf{Xuewen Luo, Ding Fan, Ruiqi Chen, Ye Cao, Xiujin Liu, Bo Yu,} \\
  \textbf{Fengze Yang, Chenxi Liu} \\
  University of Utah, Monash University}

\begin{document}
\maketitle
\begin{abstract}

As autonomous vehicles advance toward driverless mobility, understanding and responding to passenger needs and intentions becomes increasingly important in the absence of a human driver. We propose Intent2Drive, a unified framework for holistic passenger intent understanding and passenger-aligned planning. Unlike existing methods that rely on explicit commands, Intent2Drive models passenger intent as a latent cognitive state inferred from language, personal attributes, emotions, behaviors, and situational context. To support this task, we construct the Holistic Passenger Intent Dataset (HPID) with structured annotations of explicit and implicit passenger-intent cues. A Theory-of-Mind-inspired Passenger Intent Reasoner (PIR) infers a Latent Passenger State (LPS) and converts it into a planner-compatible Passenger Intent Objective (PIO). We validate the downstream utility of PIO by conditioning an existing hierarchical planning pipeline at the route and trajectory levels. Experiments demonstrate that the proposed method understands and responds to passenger needs, enabling passenger-aligned driving while maintaining competitive closed-loop planning performance.

\end{abstract}

\section{Introduction}

Autonomous driving (AD) has made remarkable progress in perception, reasoning, and planning, bringing fully driverless services such as robotaxis closer to reality \cite{chen2024end,zhao2025survey,ding2026knowdiffuser}. Yet removing the human driver also removes a critical passenger-responsive role: human drivers not only control the vehicle but also interpret and respond to passengers' needs, such as requests to stop, change destinations, or seek assistance. Existing AD systems remain largely machine-centered, optimizing predefined objectives from external sensor observations without adequately perceiving, reasoning about, or responding to in-cabin passenger intent \cite{cui2024drive}. Driverless mobility therefore calls for human-centered systems that can understand passenger needs and adapt their driving behavior accordingly \cite{wu2021human,huang2025toward,luo2025s}.

Existing human-in-the-loop AD approaches, however, provide an \textbf{underspecified representation of passenger intent} \cite{wu2023toward}. They commonly treat intent as spoken language or low-level commands, such as ``slow down,'' ``change lanes,'' or ``turn left,'' and map these inputs directly to driving actions \cite{xu2025chatmpc,ma2024lampilot}. Such representations capture explicit requests but overlook implicit passenger-related cues, including emotional and physical states, facial expressions, body posture, personal attributes, and situational context \cite{tellex2011understanding,dragan2013legibility}. Holistic intent understanding must instead infer passenger needs from both explicit expressions and implicit cues. A further challenge is to translate the inferred passenger state into structured objectives that can guide downstream driving decisions; otherwise, intent understanding remains disconnected from executable planning.

From a cognitive perspective, passenger intent understanding can be formulated as a Theory of Mind (ToM) problem \cite{wang2024theory}. Because intent is not directly observable, an autonomous system must reason from passenger expressions, behaviors, and context to the latent goals, preferences, conditions, and constraints underlying them. This perspective motivates a two-stage bridge from observable passenger cues to a structured latent state, and from that state to a planning-compatible objective that supports passenger-aligned driving.


To address these challenges, we propose Intent2Drive, a unified framework for holistic passenger intent understanding and passenger-aligned planning. We construct the Holistic Passenger Intent Dataset (HPID), which captures explicit intent and diverse implicit passenger-related cues while providing structured supervision for a Passenger Intent Reasoner (PIR). Inspired by ToM, PIR infers a Latent Passenger State (LPS) and translates it into a planner-compatible Passenger Intent Objective (PIO). We then condition a hierarchical route- and trajectory-planning pipeline on PIO to evaluate whether the inferred intent can guide executable, passenger-aligned driving behavior.

\textbf{Our contributions are threefold:}

\begin{itemize}

\item We formulate holistic passenger intent understanding and its connection to downstream planning, and propose \textbf{Intent2Drive}, a unified framework for passenger intent understanding and passenger-aligned driving while maintaining competitive closed-loop planning performance.

\item We introduce \textbf{HPID} to provide structured supervision for intent reasoning. To our knowledge, it is the first real-world dataset to jointly capture explicit intent and diverse implicit passenger-related cues.

\item A ToM-inspired \textbf{PIR} is introduced to bridge passenger intent understanding and driving by inferring a cognitive state (\textbf{LPS}) from diverse passenger-related cues and translating it into a planner-compatible \textbf{PIO}.

\end{itemize}

\section{Related Work}

\subsection{Human-in-the-Loop Autonomous Driving}


Human involvement has become increasingly important in AD, as effective human-vehicle collaboration improves user trust, acceptance, and driving experience \cite{wu2023toward, huang2025toward, ding2025drive}. Existing human-in-the-loop AD systems incorporate human guidance through reward design, preference learning, reinforcement learning from human feedback, or natural-language interaction. Early methods primarily relied on human feedback to optimize driving policies \cite{sadigh2017active, christiano2017deep}, while recent large language model (LLM)-based approaches enable users to communicate with autonomous vehicles (AVs) using natural-language instructions, preferences, and explanations \cite{xu2025chatmpc, ma2024lampilot}.


Despite these advances, existing human-in-the-loop AD systems primarily infer passenger intent from explicit user inputs, such as preference feedback, demonstrations, or natural-language instructions \cite{sadigh2017active, christiano2017deep, xu2025chatmpc, ma2024lampilot}. Consequently, latent factors underlying passenger intent are largely excluded from decision-making. However, research in human--robot interaction and cognitive science shows that passenger intent is jointly shaped by explicit language and diverse implicit passenger-related cues, including emotional and physical states, behavioral signals, body language, personal attributes, and situational context \cite{tellex2011understanding, dragan2013legibility, huang2025toward}. Therefore, relying solely on explicit inputs yields an inherently incomplete representation of passenger intent. In contrast, our work constructs a holistic representation by jointly reasoning over explicit intent and implicit passenger-related cues.

\subsection{ToM for Passenger Intent Understanding}
\begin{figure*}[t]
  \centering
  \vspace{-2mm}
  \includegraphics[width=0.82\textwidth]{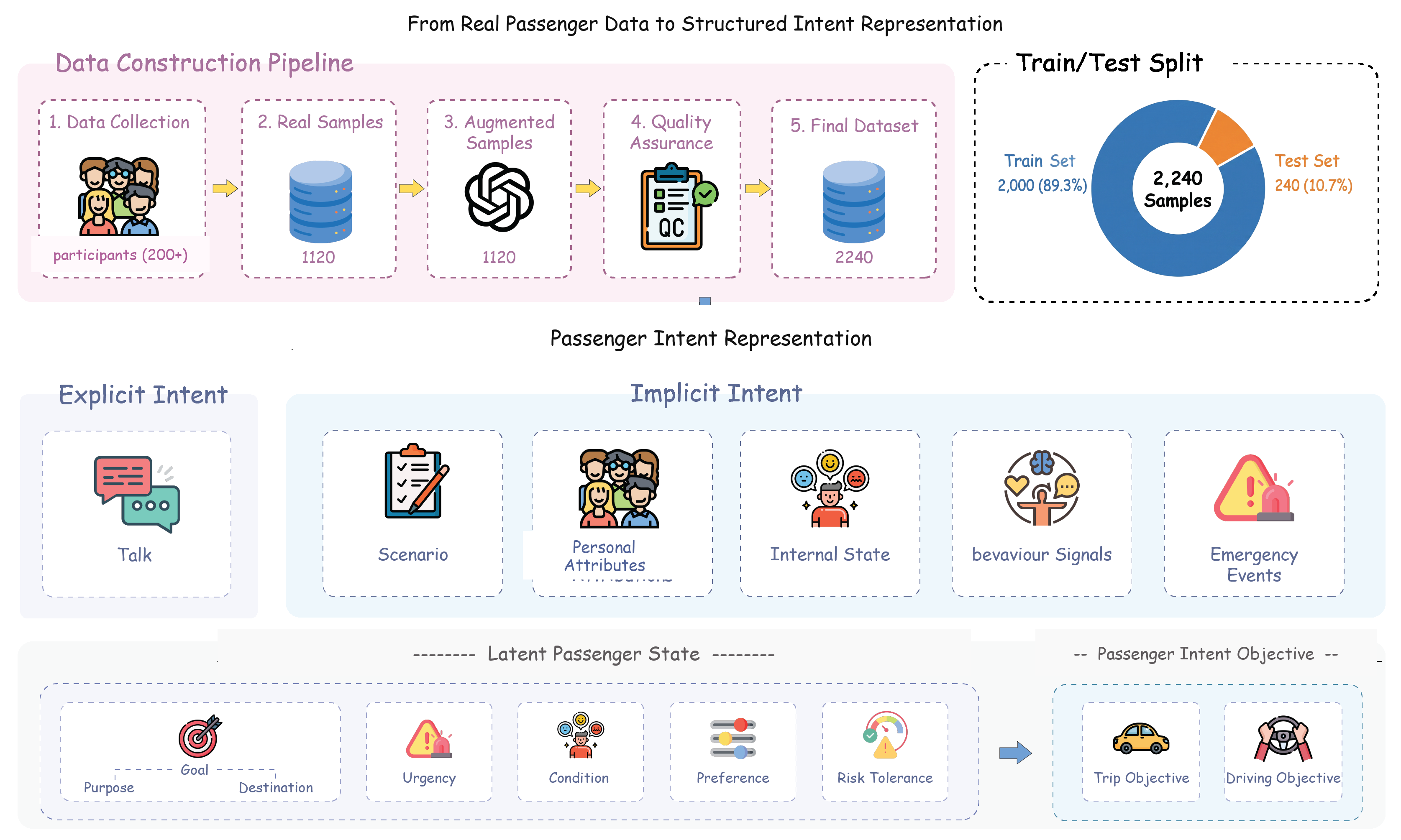}
  \caption{Pipeline for HPID Construction}
  \label{fig:hpid-construction}
  \vspace{-2mm}
\end{figure*}

ToM refers to the ability to infer unobservable mental states, such as beliefs, desires, goals, and intentions, from observable behaviors and contextual cues \cite{premack1978does, baker2011bayesian}. Rather than reasoning solely from surface-level actions or language, ToM models the latent cognitive states underlying human behavior. Inspired by this capability, computational ToM has become an important paradigm for human-centered AI \cite{rabinowitz2018machine, kosinski2024evaluating}, supporting intention inference, future action prediction, and decision-making by explicitly modeling latent mental states.


ToM-inspired reasoning has been widely applied in robotics and human-robot interaction for intention inference and cooperative decision-making \cite{tellex2011understanding, dragan2013legibility, nikolaidis2017human}. These studies demonstrate that modeling latent human states enables more effective human-aware reasoning than relying solely on observable behaviors. Building on this insight, we introduce a ToM-inspired PIR that jointly reasons over explicit intent and implicit passenger-related cues to infer an LPS, providing the cognitive foundation for downstream AD planning.

\section{Holistic Passenger Intent Dataset}

Existing human-in-the-loop AD datasets typically represent passenger intent through low-level instructions or isolated preference signals, overlooking the latent states and contextual factors underlying passenger decisions. To support holistic intent understanding, we construct HPID, illustrated in Fig.~\ref{fig:hpid-construction}, to capture a structured chain from passenger evidence to planning-oriented objectives. Each sample represents a complete trip experience and contains three components. (1) \textbf{Passenger Intent} comprises explicit expressions and implicit passenger-related cues, including scenario context, personal attributes, internal states (e.g., emotional and physical states, trust, and attention), behavioral signals, and emergency events. (2) \textbf{LPS} represents the passenger's inferred goal, urgency, condition, preference, and risk tolerance. (3) \textbf{PIO} translates passenger needs into trip and driving objectives. Thus, Passenger Intent provides observable evidence, LPS serves as an interpretable intermediate representation, and PIO provides the planning target. This hierarchy supports both intent reasoning and downstream intent-conditioned planning.

We collect participant-derived passenger-intent data from 226 adults through structured questionnaires and scenario-based interviews. Participants describe realistic driving situations, needs, behaviors, internal states, and contextual factors in natural language. The data cover commuting, time-critical travel, emergencies, comfort-oriented travel, social activities, and special passenger needs. Because each participant may contribute multiple scenarios, consistency review and noise removal yield 1,120 real samples.

To improve coverage across driving-objective classes, we use GPT-4o to construct one counterfactual counterpart for each real sample. GPT-4o preserves the scenario, changes the target driving-style class, and minimally revises one to three leaf fields in the passenger-intent record. For both real records and their counterfactual variants, GPT-4o extracts structured LPS information from passenger self-reports and their minimally revised variants, whereas passenger-reviewed PIO labels serve as ground truth. Automatic schema and consistency checks, followed by manual inspection, ensure coherence among the Scenario, Passenger Intent, LPS, and PIO. This process produces 1,120 validated synthetic samples.

HPID contains 2,240 samples, evenly divided between real and synthetic data. The test set consists exclusively of 240 real participant-derived samples and contains no synthetic data. We split the remaining 2,000 samples into 1,800 training and 200 validation samples, with synthetic samples restricted to these two splits. Thus, LPS inference and PIO construction are evaluated exclusively on real participant-derived scenarios. Full collection, annotation, and split details are provided in Appendix~\ref{app:hpid_collection} and Appendix~\ref{app:hpid_annotation}.

\begin{figure*}[t]
  \centering
  \vspace{-2mm}
  \includegraphics[width=0.81\textwidth]{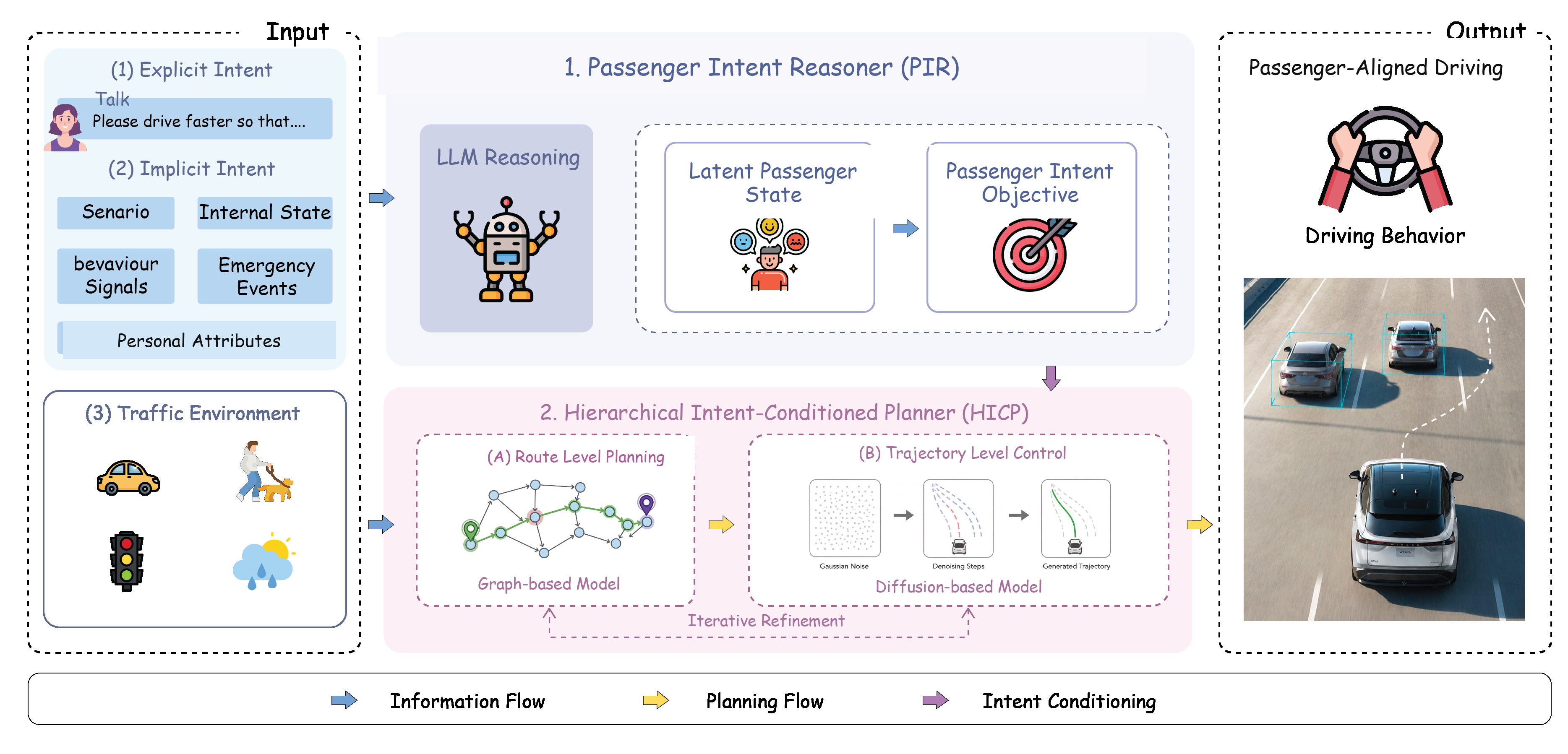}
  \caption{Framework of Intent2Drive}
  \label{framework}
  \vspace{-2mm}
\end{figure*}

\section{Methodology}

\subsection{Overview}

Intent2Drive is a human-in-the-loop AD framework designed to enable holistic passenger intent understanding and passenger-aligned driving, as illustrated in Fig.~\ref{framework}. By explicitly modeling both explicit and implicit aspects of passenger intent, the framework allows AVs to better understand passenger needs and translate them into driving behaviors that remain consistent with passenger expectations over both long and short time horizons.

The framework consists of two main components: a PIR and a route- and trajectory-planning pipeline named the Hierarchical Intent-Conditioned Planner (HICP). PIR, which serves as the core component of Intent2Drive, is built upon an LLM fine-tuned on HPID. Inspired by ToM reasoning, it infers an LPS, which summarizes the latent factors underlying passenger intent, including the passenger's goal, urgency, condition, preference, and risk tolerance. The inferred LPS is subsequently transformed into a PIO, a planning-oriented abstraction that bridges passenger intent understanding and downstream decision-making. Conditioned on the PIO, HICP serves as a downstream application that translates the inferred intent into executable driving behaviors. Specifically, it incorporates the PIO into both route-level planning and trajectory-level control, enabling decisions that remain consistent with the passenger's intent across different planning horizons. 

\subsection{Passenger Intent Reasoner}

\paragraph{Passenger Intent Representation Input}

PIR takes as input the holistic passenger intent representation from HPID. It consists of explicit intent and implicit passenger-related information, with the latter covering scenario context, personal attributes, internal states, behavioral signals, and emergency events.

Formally, the passenger intent input is represented as
\begin{equation}
\mathbf{H}=\left(x^{\mathrm{exp}},x^{\mathrm{imp}}\right),
\end{equation}
where $x^{\mathrm{exp}}$ and $x^{\mathrm{imp}}$ denote explicit intent and implicit passenger-related information, respectively. PIR uses both as observable evidence for latent-state inference.

\paragraph{Latent Passenger State Reasoning}

Passenger intent is inherently latent and cannot be directly observed. According to ToM, understanding a passenger's intent requires reasoning about the underlying mental states that give rise to observable language, expressions, and behaviors. Motivated by this perspective, we formulate passenger intent understanding as a latent mental-state inference problem. Instead of directly mapping observable passenger intent to planning objectives, PIR first infers an intermediate LPS, which serves as a cognitive bridge between observable passenger intent and downstream planning.

Rather than using an uninterpretable embedding, we represent LPS explicitly according to two design goals: (1) \textbf{cognitive coverage} of the key latent factors explaining passenger intent and (2) \textbf{planning relevance} for deriving objectives without directly encoding planning decisions. Accordingly,

\begin{equation}
m=(g,u,c,p,\rho), \qquad g=(g^{\mathrm{pur}},g^{\mathrm{dst}}),
\end{equation}

where $g^{\mathrm{pur}}$ and $g^{\mathrm{dst}}$ denote trip purpose and destination, $u$ is urgency, $c$ is the dominant emotional or physical condition, $p$ is travel preference, and $\rho$ is risk tolerance. These factors compactly describe the passenger's latent decision state without encoding a planning action.

Given the observable passenger intent input, PIR performs ToM reasoning to infer the corresponding latent passenger state,

\begin{equation}
m=\mathcal{R}\left(x^{\mathrm{exp}},x^{\mathrm{imp}}\right),
\end{equation}

where $\mathcal{R}$ denotes the conceptual LPS inference function. Unlike direct instruction following, this formulation exposes a structured latent state before constructing planning objectives.

\paragraph{Passenger Intent Objective Construction}

Although the inferred LPS provides an interpretable representation of the passenger's latent mental state, it is not directly suitable for AD planning. The LPS describes why a passenger behaves in a particular way by capturing latent cognitive factors, whereas downstream planners require explicit optimization objectives rather than mental-state representations. Therefore, a direct coupling between LPS and motion planning is neither intuitive nor efficient.

To bridge this gap, we introduce a PIO, which serves as a planning-oriented abstraction of the inferred LPS. Unlike the LPS, which focuses on understanding passenger cognition, the PIO specifies how the AV should respond to the inferred passenger intent. This decoupling separates passenger intent understanding from downstream planning, allowing the reasoning process to remain cognitively interpretable while providing a planner-compatible representation.

Formally, the PIO is defined as

\begin{equation}
z=(o,d),
\end{equation}

where $o$ is the destination-level trip objective derived from $g$, and $d\in\{\texttt{Aggressive},\texttt{Standard},\texttt{Conservative}\}$ is the driving objective derived from $(u,c,p,\rho)$.

The PIO is constructed from the inferred latent passenger state through

\begin{equation}
z=\mathcal{F}_{\mathrm{PIO}}(m),
\end{equation}

where $\mathcal{F}_{\mathrm{PIO}}$ denotes the conceptual PIO construction function. This decomposition separates passenger understanding from planning abstraction and provides a compact interface for intent-conditioned planning.

\paragraph{Training Strategy}

PIR is trained by supervised fine-tuning (SFT) on HPID using a Qwen3-4B backbone and Low-Rank Adaptation (LoRA). The functions $\mathcal{R}$ and $\mathcal{F}_{\mathrm{PIO}}$ describe a conceptual decomposition: in implementation, one model jointly generates LPS and PIO in a single autoregressive response, with PIO tokens conditioned on the preceding LPS tokens. Thus, LPS is an explicit intermediate prediction rather than a free-text reasoning trace. Training and inference details are provided in Appendix~\ref{app:pir_training_details}.

\subsection{Hierarchical Intent-Conditioned Planner}

\paragraph{Problem Formulation}

To evaluate whether PIO provides an actionable interface rather than only an accurate structured prediction, we instantiate HICP as a downstream application. HICP is not intended as a standalone planning contribution; it serves as a testbed that injects the inferred PIO into established route-planning and trajectory-generation stages. Given an environment state $e$, containing static map information and dynamic traffic observations, and a PIO $z$, the application produces intent-conditioned plans.

Formally, the planner is defined as

\begin{equation}
(r,\tau)=\mathcal{P}(e,z),
\end{equation}

where $r$ and $\tau$ denote the route- and trajectory-level plans. Conditioning both stages on PIO allows us to test whether the proposed representation induces consistent behavioral changes across planning horizons while retaining feasibility and general planning performance.

The two PIO components have complementary roles across the hierarchy. The trip objective $o$ specifies the destination-level target and primarily guides route selection, whereas the driving objective $d$ controls how that target is pursued at both levels. In particular, $d$ distinguishes progress-oriented, balanced, and conservative behavior without replacing the planner's environmental and safety constraints.

\paragraph{Intent-Conditioned Route Planning}

The road network is represented as a directed graph
\(
\mathcal{G}=(\mathcal{V},\mathcal{E})
\),
where nodes denote lane segments and edges represent feasible transitions. Route planning is formulated as sequential decision-making over the graph to generate a route
\(
r=\{v_0,\ldots,v_T\}.
\)

A base routing policy is first learned from expert demonstrations using supervised learning,

\begin{equation}
\mathcal{L}_{\mathrm{route}}
=
-
\mathbb{E}_{r^\ast}
\sum_t
\log p_\theta(v_{t+1}^\ast|v_t^\ast,e),
\end{equation}

which captures fundamental navigation patterns and topological constraints.

The routing policy is subsequently optimized using reinforcement learning, with $z$ supplied as a conditioning variable at each routing step,

\begin{equation}
\nabla_\theta J(\theta)
=
\mathbb{E}_{e,z,r}
\left[
R_z(r)
\nabla_\theta
\log p_\theta(r|e,z)
\right],
\end{equation}

where the expectation covers context pairs $(e,z)$ and routes $r\sim p_\theta(\cdot|e,z)$. We decompose the intent-conditioned reward as

\begin{equation}
R_z(r)=R_{\mathrm{trip}}(r;o)+R_{\mathrm{drive}}(r;d),
\end{equation}

where $R_{\mathrm{trip}}$ rewards reaching the trip objective and $R_{\mathrm{drive}}$ ranks feasible routes according to the requested driving behavior. During inference, the planner generates

\begin{equation}
r=\mathcal{P}_r(e,z),
\end{equation}

which serves as the global guidance for trajectory generation.

\paragraph{Intent-Conditioned Trajectory Generation}

Given the environment state $e$, planned route $r$, and PIO $z$, we formulate trajectory planning as conditional generation of a short-horizon trajectory. The route supplies navigation guidance, while PIO specifies the desired behavior.

We adopt an $x^{(0)}$-prediction diffusion model. Given ground-truth trajectory $x^{(0)}$, the forward process samples $t$ and $\epsilon\sim\mathcal{N}(0,I)$ and constructs

\begin{equation}
x^{(t)}=\sqrt{\bar{\alpha}_t}x^{(0)}+\sqrt{1-\bar{\alpha}_t}\epsilon,
\end{equation}

where $\bar{\alpha}_t$ is the cumulative noise schedule. The denoiser is trained by

\begin{equation}
\mathcal{L}_\theta
=
\mathbb{E}_{x^{(0)},t,\epsilon}
\left[
\left\|
f_\theta(x^{(t)},t,C_e,C_r,z)
-
x^{(0)}
\right\|^2
\right],
\end{equation}

where $C_e$ and $C_r$ denote the encoded environment and route context, respectively.

Here, $C_e$ preserves scene and traffic constraints, $C_r$ carries the global guidance associated with $o$, and $d$ provides the local behavioral condition through $z$. This separation allows different driving styles to be generated for the same route and environment while leaving the underlying scene constraints unchanged.

During inference, iterative reverse diffusion conditioned on $(C_e,C_r,z)$ yields

\begin{equation}
\tau=\mathcal{P}_\tau(e,r,z),
\end{equation}

producing intent-conditioned trajectory candidates that are subsequently handled by the pipeline's existing feasibility and safety constraints.

\section{Experiment}

\subsection{Experimental Setup}

We evaluate PIR on held-out HPID samples in terms of LPS inference, PIO construction, the contributions of LPS reasoning and implicit cues, and qualitative ToM reasoning. We then evaluate Intent2Drive on the official nuPlan simulator \cite{caesar2021nuplan} under non-reactive and reactive closed-loop settings to assess intent-conditioned planning and general planning performance.

\subsection{Passenger Intent Understanding}

Unlike instruction-following approaches that predict planning objectives directly from language, PIR first infers an LPS and then constructs a PIO. We evaluate (1) LPS inference and PIO construction, (2) the contribution of LPS reasoning, (3) the contribution of implicit passenger-related cues, and (4) the interpretability of this ToM reasoning process through qualitative analysis.

All reported PIR accuracies are means over three independent runs. Categorical fields (urgency, risk tolerance, and driving objective) use normalized exact match, while GPT-4o judges semantic equivalence for open-ended fields using an evaluation prompt distinct from the annotation prompts. Goal accuracy is the macro-average of purpose and destination accuracy; aggregate LPS accuracy in the ablations is the macro-average over the five LPS attributes.

\paragraph{LPS Inference and PIO Construction}

\begin{table}[H]
\centering
\footnotesize
\caption{Performance of LPS inference (\%).}
\label{tab:lps}
\begin{tabular}{lcc}
\toprule
LPS Attribute & Pretrained Qwen3-4B (\%) & Ours (\%) \\
\midrule
Goal       & 5.63  & 69.79 \\
Urgency    & 60.42 & 85.00\\
Condition  & 17.50 & 76.67 \\
Preference & 52.92 & 84.58 \\
Risk Tolerance & 9.17  & 90.42 \\
\bottomrule
\end{tabular}
\end{table}

\begin{table}[H]
\centering
\footnotesize
\caption{Performance of PIO construction (\%).}
\label{tab:pio}
\begin{tabular}{lcc}
\toprule
PIO Component & Pretrained Qwen3-4B (\%) & Ours (\%) \\
\midrule
Trip Objective    & 37.50 & 81.25 \\
Driving Objective & 60.83 & 85.42 \\
\bottomrule
\end{tabular}
\end{table}

Tables~\ref{tab:lps} and~\ref{tab:pio} show that PIR consistently outperforms pretrained Qwen3-4B across both reasoning stages. LPS accuracy rises from 5.63--60.42\% to 69.79--90.42\%, with especially large gains on goal, condition, preference, and risk tolerance, which require integrating language with contextual evidence. PIR also improves trip- and driving-objective accuracy from 37.50\% and 60.83\% to 81.25\% and 85.42\%, respectively. The particularly large trip-objective gain indicates that destination-level objectives benefit from explicitly reasoning about the passenger's underlying goal and condition. Together, these results show that PIR reliably predicts the structured LPS annotations and translates them into planner-compatible objectives. Compared with larger general-purpose models, HPID-supervised Qwen3-4B achieves the best result on six of seven fields, demonstrating that the compact model remains highly competitive (Appendix Table~\ref{tab:additional_large_model_pio}). HPID-supervised SFT also produces substantial improvements at every Qwen3 scale, showing that its benefit is consistent across model capacities (Appendix Table~\ref{tab:qwen_scaling_hpid}).

\paragraph{Effect of LPS Reasoning}

To isolate the contribution of LPS reasoning, we compare PIR with a direct-PIO baseline that maps the same holistic input directly to trip and driving objectives. Both use the same Qwen3-4B backbone, data, and optimization; they differ only in whether the LPS is explicitly modeled.

\begin{table}[t]
\centering
\footnotesize
\caption{Effect of LPS reasoning on PIO construction (\%).}
\label{tab:lps_ablation}
\begin{tabular}{lccc}
\toprule
Method & LPS & Trip Obj. & Driving Obj. \\
\midrule
Direct PIO & No  & 79.17 & 81.25 \\
PIR (Ours) & Yes & 81.25 & 85.42 \\
\bottomrule
\end{tabular}
\end{table}

Table~\ref{tab:lps_ablation} uses the same held-out test set, metrics, backbone, and optimization protocol as Table~\ref{tab:pio}. LPS reasoning improves trip- and driving-objective accuracy by 2.08 and 4.17 points, respectively. The larger driving-objective gain reflects its dependence on interacting factors such as urgency, comfort preference, and risk tolerance. These descriptive gains indicate that the structured LPS provides a useful intermediate representation rather than merely an auxiliary output.

\paragraph{Contribution of Implicit Passenger Intent}

To investigate the importance of implicit passenger-related information, we conduct an ablation study by progressively introducing different intent components into PIR.

\begin{table}[t]
\centering
\footnotesize
\caption{Cumulative contribution of passenger intent components to LPS inference and PIO construction. PIO is the macro-average accuracy over trip and driving objectives.}
\label{tab:ablation}
\begin{tabular}{lcc}
\toprule
Cumulative Input & LPS (\%) & PIO (\%) \\
\midrule
Explicit Intent Only (Talk) & 46.53 & 51.25 \\
+ Personal Attributes & 50.35 & 53.96 \\
+ Internal States & 61.32 & 57.50 \\
+ Behavioral Signals & 62.15 & 58.54 \\
+ Emergency Event & 63.13 & 58.96 \\
+ Scenario & 81.29 & 83.34 \\
\bottomrule
\end{tabular}
\end{table}

Table~\ref{tab:ablation} shows cumulative gains as the input components are added sequentially in the displayed order. Explicit intent alone reaches 46.53\% LPS and 51.25\% PIO accuracy. Personal attributes provide passenger-specific priors, internal states add affective and physical evidence, and behavioral signals and emergency events provide further incremental gains. Adding scenario context last produces the largest step, raising LPS accuracy to 81.29\% and PIO accuracy to 83.34\%; the latter averages the 81.25\% trip-objective and 85.42\% driving-objective accuracies in Table~\ref{tab:pio}. Thus, under this cumulative ordering, every added component contributes to holistic inference.

\paragraph{Qualitative Analysis}

Appendix Tables~\ref{tab:pir_case_motion_sickness}--\ref{tab:pir_case_school_aggressive} qualitatively assess the interpretability of PIR's ToM reasoning across six diverse contexts, spanning physical discomfort, caregiving, routine travel, environmental concerns, and time-critical trips. Their Conservative, Standard, and Aggressive outcomes show how urgency, condition, preference, and risk tolerance jointly shape planning objectives. By exposing the LPS-to-PIO chain, the cases make the intermediate reasoning traceable rather than directly mapping language to actions.

\subsection{Closed-Loop Planning Performance}

Because nuPlan lacks in-cabin passenger information, we pair each driving scenario with a manually curated HPID context without altering its map or dynamic traffic observations. PIR constructs $z=(o,d)$, which conditions the route policy and reward $R_z$ as well as the trajectory generator together with $(C_e,C_r)$. We assess the resulting behavior and general closed-loop performance.

Appendix Fig.~\ref{fig:pio_planning_trajectories} shows that different PIOs produce distinct, feasible trajectories in the same driving context.

We further compare Intent2Drive with expert, rule-based, learning-based, diffusion-based, and LLM-assisted planners using the official aggregate nuPlan closed-loop score (higher is better) on Val14 and Test14. Non-reactive (NR) evaluation replays logged agent behavior, whereas reactive (R) evaluation allows other agents to respond to the ego vehicle.

\begin{table}[t]
\centering
\footnotesize
\caption{Closed-loop planning performance on the nuPlan benchmark.}
\label{tab:nuplan_closed_loop}
\setlength{\tabcolsep}{2.2pt}

\begin{tabular}{p{4.4cm}cccc}
\toprule
\textbf{Method}
& \multicolumn{2}{c}{\textbf{Val14}}
& \multicolumn{2}{c}{\textbf{Test14}} \\
\cmidrule(lr){2-3} \cmidrule(lr){4-5}
& NR & R & NR & R \\
\midrule
Log-replay (Expert) & 93.53 & 80.32 & 94.03 & 75.86 \\
IDM \cite{treiber2000congested} & 75.60 & 77.33 & 70.39 & 74.42 \\
PDM-Closed \cite{caesar2021nuplan} & 92.84 & 92.12 & 90.05 & 91.63 \\
PDM-Hybrid \cite{caesar2021nuplan} & 92.77 & 92.11 & 90.10 & 91.28 \\
UrbanDriver \cite{scheel2022urban} & 68.57 & 64.11 & 51.83 & 67.15 \\
PlanTF \cite{cheng2024rethinking} & 84.27 & 76.95 & 85.62 & 79.58 \\
GameFormer \cite{huang2023gameformer} & 79.94 & 79.78 & 83.88 & 82.05 \\
PLUTO \cite{cheng2024pluto} & 92.88 & 76.88 & 92.23 & 90.29 \\
Diffusion Planner \cite{zheng2025diffusion} & 89.87 & 82.80 & 89.19 & 82.93 \\
Flow Planner \cite{tan2025flow} & 90.43 & 83.31 & 89.88 & 82.93 \\
LLM-ASSIST \cite{sharan2023llm} & 93.00 & 92.20 & -- & -- \\
\midrule
\textbf{Intent2Drive}
& 93.02 & 84.05
& 87.20 & 81.35 \\
\bottomrule
\end{tabular}
\end{table}

Table~\ref{tab:nuplan_closed_loop} shows that Intent2Drive achieves 93.02/84.05 (NR/R) on Val14 and 87.20/81.35 on Test14, remaining competitive with learning- and diffusion-based planners. Performance is stronger in the non-reactive setting, while the reactive scores reflect the additional difficulty of accounting for other agents' responses. Together with Appendix Fig.~\ref{fig:pio_planning_trajectories}, these results provide complementary evidence: the benchmark scores establish the closed-loop feasibility of the combined system, whereas the controlled trajectories show that PIO changes behavior in the intended direction. Thus, PIO supports behavioral differentiation while the resulting system retains competitive planning capability.

\section{Conclusion}

We presented Intent2Drive, a human-in-the-loop AD framework that jointly reasons over explicit intent and implicit passenger-related cues. To support this task, we constructed HPID with structured supervision spanning Passenger Intent, LPS, and PIO. The ToM-inspired PIR infers a structured LPS and translates it into a planner-compatible PIO. Experiments on held-out real participant-derived samples show consistent improvements in LPS and PIO prediction across model scales, while ablations and qualitative cases demonstrate how passenger state and context inform the resulting objectives. Conditioning a route- and trajectory-planning pipeline on PIO produces behaviorally distinct trajectories in the intended direction, while the resulting system remains competitive in closed-loop planning. Together, these findings establish structured intent reasoning as an actionable interface between passenger understanding and AD planning.

\section{Limitations}

Intent2Drive has two main limitations. First, HPID represents passenger evidence as textual, structured records rather than synchronized raw in-cabin observations. PIR therefore cannot model temporal patterns in gaze, prosody, facial expression, or body movement. Moreover, because LPS annotations are extracted from passenger self-reports using GPT-4o and then quality-checked, the reported results measure agreement with structured proxy labels rather than direct recovery of unobservable mental states. Future work should incorporate synchronized video, audio, and physiological signals with stronger human annotation and validation. Second, evidence for passenger alignment is currently limited to offline dataset evaluation and nuPlan simulation with manually paired passenger contexts. We have not conducted studies with human passengers or real-vehicle tests to determine whether the generated behaviors are perceived as aligned, comfortable, trustworthy, and safe. Controlled user studies and real-world passenger evaluations are therefore necessary before deployment-oriented claims can be made.

\clearpage


\bibliography{custom}
\clearpage

\appendix
\raggedbottom

\section{Appendix}

\subsection{Ethical Considerations}
\label{app:ethical_considerations}

Intent2Drive presents several potential ethical and societal risks. First, errors in latent passenger-state inference or PIO construction could lead the AV to misinterpret a passenger's intent and generate inappropriate driving behaviors, with potentially serious consequences in safety-critical situations. The current system is therefore intended solely as a research prototype. It should not replace established perception, planning, or safety mechanisms, nor be deployed without extensive real-world validation, uncertainty estimation, fail-safe constraints, and appropriate human oversight.

Second, although HPID includes participants from diverse backgrounds, substantial variability remains in how individuals interpret the same high-level driving preference. For example, two participants who both prefer an "aggressive" driving style may nonetheless disagree on acceptable following distances, acceleration patterns, lane-changing frequency, or overtaking behaviors. Consequently, the current coarse-grained preference taxonomy cannot fully capture fine-grained individual driving preferences, and models trained on HPID may produce behaviors that align better with some passengers than others. Future work will enrich the dataset by modeling personalized fine-grained preferences conditioned on shared high-level preference categories, enabling the system to preserve general behavioral consistency while adapting to individual driving expectations.

Third, in-cabin language, behavioral signals, and personal attributes are inherently privacy-sensitive. In the current study, all participants were adults, participation was voluntary with informed consent, and the questionnaire and interview data were anonymized without retaining direct personal identifiers. Nevertheless, any future data collection or public release should continue to minimize retained personal information, restrict access to raw records, and clearly communicate the intended use of the data to participants.

Finally, premature deployment of passenger-intent-aware driving systems without sufficient validation could disproportionately affect vulnerable passengers if intent inference fails or personalized behaviors are not well calibrated. These risks highlight the importance of transparent auditing, fairness evaluation across diverse user groups, and rigorous testing under realistic interactive driving scenarios before any real-world deployment.

\subsection{Case Examples}

\begin{table*}[!t]
\centering
\small
\setlength{\tabcolsep}{6pt}
\renewcommand{\arraystretch}{1.1}

\begin{tabular}{@{}p{0.96\textwidth}@{}}
\toprule

\begin{minipage}[t]{\linewidth}
\textbf{Case 1: Motion Sickness}\\[2pt]

\textbf{Explicit Intent:}\\
Talk: Ugh, I feel so nauseous right now\ldots{} I wish the car would slow down a little or make a stop\ldots{}\\[3pt]

\textbf{Implicit Intent:}\\
\textit{Scenario:} A female manager in her late 30s is experiencing severe motion sickness, feeling nauseous and dizzy, while riding in an AV with her partner.\\
\textit{Age Range:} 35--44\\
\textit{Gender:} Female\\
\textit{Occupation:} Manager\\
\textit{Passenger Frequency:} Almost never\\
\textit{Driving Experience:} More than 10 years\\
\textit{AV Experience:} No, but I have tried advanced driver-assistance systems, e.g., Tesla Autopilot\\
\textit{AV Trust Baseline:} Low\\
\textit{Passenger Monitoring Tendency:} Very much --- I watch the road constantly and often want to give directions\\
\textit{Risk Attitude:} Somewhat cautious --- I prefer the safer option but can take calculated risks\\
\textit{Physical Condition Baseline:} Back pain or joint issues\\
\textit{Social Context:} Partner\\[3pt]

\textbf{Passenger State:}\\
\textit{Emotion:} Physically uncomfortable, vulnerable, irritable / easily annoyed, worried\\
\textit{Physical State:} Nauseous, dizzy, upset stomach, lightheaded\\
\textit{Comfort State:} Uncomfortable\\
\textit{Trust State:} Low trust in AV\\
\textit{Attention State:} High monitoring\\
\textit{Cognitive Load:} High\\[3pt]

\textbf{Observable Signals:}\\
\textit{Facial Expression:} None\\
\textit{Body Posture:} Hunched over or curled up; leaning back with eyes closed; holding stomach or head; resting head against window or seat; sitting very still\\
\textit{Gaze Behavior:} None\\
\textit{Hand Motion:} None\\
\textit{Behavior in Ride:} Frequently adjusting position to find comfort\\[3pt]

\textbf{Emergency Event:}\\
\textit{Type:} Motion sickness\\
\textit{Description:} Severe nausea and dizziness due to car movement\\
\textit{Severity:} High\\
\textit{Trigger Source:} Car movement\\
\textit{Planning Impact:} Passenger prioritizes physical comfort over punctuality\\[3pt]

\textbf{LPS:}\\
\textit{Goal:} Receive relief from motion sickness\\
\textit{Destination:} Safe roadside location\\
\textit{Urgency:} High\\
\textit{Condition:} Unwell\\
\textit{Preference:} Comfort\\
\textit{Risk Tolerance:} Low\\[3pt]

\textbf{PIO:}\\
\textit{Trip Objective:} \textbf{Safe Roadside Location}\\
\textit{Driving Objective:} \textbf{Conservative}\\
\end{minipage}

\\
\bottomrule
\end{tabular}

\caption{
Qualitative example of PIR for a motion-sickness passenger.
}
\label{tab:pir_case_motion_sickness}
\end{table*}

\begin{table*}[!t]
\centering
\small
\setlength{\tabcolsep}{6pt}
\renewcommand{\arraystretch}{1.1}

\begin{tabular}{@{}p{0.96\textwidth}@{}}
\toprule

\begin{minipage}[t]{\linewidth}
\textbf{Case 2: Urgent Hospital Trip with Baby}\\[2pt]

\textbf{Explicit Intent:}\\
Talk: Please drive as smoothly as possible, I'm holding my one-month-old baby. Every little bump makes me worry he'll wake up or cry. I need this ride to be absolutely gentle.\\[3pt]

\textbf{Implicit Intent:}\\
\textit{Scenario:} A mother, holding her one-month-old baby, urgently needs to reach the hospital due to a family accident, while feeling anxious and stressed.\\
\textit{Age Range:} 25--34\\
\textit{Gender:} Female\\
\textit{Occupation:} Currently on maternity leave from my job as an administrative assistant.\\
\textit{Passenger Frequency:} A few times a week\\
\textit{Driving Experience:} More than 10 years\\
\textit{AV Experience:} No, never\\
\textit{AV Trust Baseline:} Low\\
\textit{Passenger Monitoring Tendency:} Very much --- I watch the road constantly and often want to give directions\\
\textit{Risk Attitude:} Very risk-averse --- I avoid risks whenever possible\\
\textit{Physical Condition Baseline:} Anxiety disorder or panic attacks\\
\textit{Social Context:} Husband\\[3pt]

\textbf{Passenger State:}\\
\textit{Emotion:} Anxious, stressed, panicked\\
\textit{Physical State:} None\\
\textit{Comfort State:} Uncomfortable due to anxiety and stress\\
\textit{Trust State:} Low trust in AV\\
\textit{Attention State:} Highly attentive to road conditions\\
\textit{Cognitive Load:} High due to stress and urgency\\[3pt]

\textbf{Observable Signals:}\\
\textit{Facial Expression:} None\\
\textit{Body Posture:} Leaning forward in my seat; sitting rigidly / tensely\\
\textit{Gaze Behavior:} Looking out the window at traffic anxiously\\
\textit{Hand Motion:} Tapping fingers, fidgeting, or bouncing leg\\
\textit{Behavior in Ride:} Sighing or taking deep breaths\\[3pt]

\textbf{Emergency Event:}\\
\textit{Type:} None\\
\textit{Description:} None\\
\textit{Severity:} None\\
\textit{Trigger Source:} None\\
\textit{Planning Impact:} None\\[3pt]

\textbf{LPS:}\\
\textit{Goal:} Reach the hospital due to a family accident\\
\textit{Destination:} Hospital\\
\textit{Urgency:} High\\
\textit{Condition:} Anxious\\
\textit{Preference:} Comfort\\
\textit{Risk Tolerance:} Low\\[3pt]

\textbf{PIO:}\\
\textit{Trip Objective:} \textbf{Hospital}\\
\textit{Driving Objective:} \textbf{Conservative}\\
\end{minipage}

\\
\bottomrule
\end{tabular}

\caption{
Qualitative example of PIR for an urgent hospital trip with an infant passenger.
}
\label{tab:pir_case_hospital_baby}
\end{table*}

\begin{table*}[!t]
\centering
\small
\setlength{\tabcolsep}{6pt}
\renewcommand{\arraystretch}{1.1}
\begin{tabular}{@{}p{0.96\textwidth}@{}}
\toprule
\begin{minipage}[t]{\linewidth}
\textbf{Case 3: Restful Commute Home After Physical Work}\\[2pt]

\textbf{Explicit Intent:}\\
Talk: Please drive smoothly and safely. I'm very tired after a long day of work and just want to relax.\\[3pt]

\textbf{Implicit Intent:}\\
\textit{Scenario:} A housekeeper in her late 40s is riding alone in an AV after a physically demanding workday, seeking a smooth and restful trip home because of exhaustion and muscle aches.\\
\textit{Age Range:} 45--54\\
\textit{Gender:} Female\\
\textit{Occupation:} Housekeeper\\
\textit{Passenger Frequency:} A few times a week\\
\textit{Driving Experience:} More than 10 years\\
\textit{AV Experience:} No, never\\
\textit{AV Trust Baseline:} Low\\
\textit{Passenger Monitoring Tendency:} A little --- mostly relaxed but paying some attention\\
\textit{Risk Attitude:} Somewhat cautious --- prefers the safer option but accepts calculated risks\\
\textit{Physical Condition Baseline:} Back pain or joint issues\\
\textit{Social Context:} Alone\\[3pt]

\textbf{Passenger State:}\\
\textit{Emotion:} Tired, exhausted\\
\textit{Physical State:} Muscle aches, back pain\\
\textit{Comfort State:} Seeking comfort\\
\textit{Trust State:} Low trust in AV\\
\textit{Attention State:} Low attention\\
\textit{Cognitive Load:} Low\\[3pt]

\textbf{Observable Signals:}\\
\textit{Facial Expression:} Tired\\
\textit{Body Posture:} Relaxed\\
\textit{Gaze Behavior:} Eyes closed\\
\textit{Hand Motion:} None\\
\textit{Behavior in Ride:} Quiet, resting\\[3pt]

\textbf{Emergency Event:}\\
\textit{Type:} None\\
\textit{Description:} None\\
\textit{Severity:} None\\
\textit{Trigger Source:} None\\
\textit{Planning Impact:} None\\[3pt]

\textbf{LPS:}\\
\textit{Goal:} Return home safely\\
\textit{Destination:} Home\\
\textit{Urgency:} Low\\
\textit{Condition:} Exhausted with muscle and back pain\\
\textit{Preference:} Balanced\\
\textit{Risk Tolerance:} Medium\\[3pt]

\textbf{PIO:}\\
\textit{Trip Objective:} \textbf{Home}\\
\textit{Driving Objective:} \textbf{Standard}\\
\end{minipage}
\\
\bottomrule
\end{tabular}
\caption{Qualitative example of PIR for a low-urgency commute balancing comfort and routine progress.}
\label{tab:pir_case_commute_standard}
\end{table*}

\begin{table*}[!t]
\centering
\small
\setlength{\tabcolsep}{6pt}
\renewcommand{\arraystretch}{1.1}
\begin{tabular}{@{}p{0.96\textwidth}@{}}
\toprule
\begin{minipage}[t]{\linewidth}
\textbf{Case 4: Eco-Conscious Trip to a Coastal Wetland}\\[2pt]

\textbf{Explicit Intent:}\\
Talk: Please take the most eco-friendly route to the wetland reserve and watch out for wildlife near the road.\\[3pt]

\textbf{Implicit Intent:}\\
\textit{Scenario:} A male environmental volunteer in his late 20s is riding alone to a coastal wetland reserve for an early-morning bird survey in light fog. He is calm, focused, and attentive to roadside wildlife.\\
\textit{Age Range:} 25--34\\
\textit{Gender:} Male\\
\textit{Occupation:} Full-time environmental conservation volunteer\\
\textit{Passenger Frequency:} A few times a month\\
\textit{Driving Experience:} 5--10 years\\
\textit{AV Experience:} No, never\\
\textit{AV Trust Baseline:} Moderate\\
\textit{Passenger Monitoring Tendency:} A little --- mostly relaxed but paying some attention\\
\textit{Risk Attitude:} Moderate --- depends on the situation\\
\textit{Physical Condition Baseline:} None\\
\textit{Social Context:} Alone; meeting other volunteers at the destination\\[3pt]

\textbf{Passenger State:}\\
\textit{Emotion:} Calm, focused, emotionally fulfilled\\
\textit{Physical State:} Physically alert\\
\textit{Comfort State:} Comfortable\\
\textit{Trust State:} Moderate trust in AV\\
\textit{Attention State:} Moderately attentive\\
\textit{Cognitive Load:} Low\\[3pt]

\textbf{Observable Signals:}\\
\textit{Facial Expression:} Calm\\
\textit{Body Posture:} Relaxed\\
\textit{Gaze Behavior:} Focused on the surroundings\\
\textit{Hand Motion:} Holding binoculars and a field notebook\\
\textit{Behavior in Ride:} Observing wildlife\\[3pt]

\textbf{Emergency Event:}\\
\textit{Type:} None\\
\textit{Description:} None\\
\textit{Severity:} None\\
\textit{Trigger Source:} None\\
\textit{Planning Impact:} None\\[3pt]

\textbf{LPS:}\\
\textit{Goal:} Conduct a bird survey\\
\textit{Destination:} Coastal wetland reserve\\
\textit{Urgency:} Low\\
\textit{Condition:} Calm\\
\textit{Preference:} Balanced\\
\textit{Risk Tolerance:} Medium\\[3pt]

\textbf{PIO:}\\
\textit{Trip Objective:} \textbf{Coastal Wetland Reserve}\\
\textit{Driving Objective:} \textbf{Standard}\\
\end{minipage}
\\
\bottomrule
\end{tabular}
\caption{Qualitative example of PIR for a low-urgency, eco-conscious trip with balanced driving needs.}
\label{tab:pir_case_wetland_standard}
\end{table*}

\begin{table*}[!t]
\centering
\small
\setlength{\tabcolsep}{6pt}
\renewcommand{\arraystretch}{1.1}
\begin{tabular}{@{}p{0.96\textwidth}@{}}
\toprule
\begin{minipage}[t]{\linewidth}
\textbf{Case 5: Last-Minute Business Flight}\\[2pt]

\textbf{Explicit Intent:}\\
Talk: Please take the fastest route to the airport. I can handle some bumps if it saves time. Keep us moving.\\[3pt]

\textbf{Implicit Intent:}\\
\textit{Scenario:} A middle-aged executive is rushing to the airport for a last-minute business trip. He is slightly stressed but focused, while his assistant reviews presentation slides beside him.\\
\textit{Age Range:} 45--54\\
\textit{Gender:} Male\\
\textit{Occupation:} Middle-level executive\\
\textit{Passenger Frequency:} A few times a week\\
\textit{Driving Experience:} More than 10 years\\
\textit{AV Experience:} No, but has used advanced driver-assistance systems\\
\textit{AV Trust Baseline:} Moderate\\
\textit{Passenger Monitoring Tendency:} Somewhat\\
\textit{Risk Attitude:} Moderate\\
\textit{Physical Condition Baseline:} None\\
\textit{Social Context:} Assistant\\[3pt]

\textbf{Passenger State:}\\
\textit{Emotion:} Slightly stressed, focused, determined\\
\textit{Physical State:} Physically tired\\
\textit{Comfort State:} Willing to sacrifice comfort for speed\\
\textit{Trust State:} Moderate trust in AV\\
\textit{Attention State:} Focused on time constraints\\
\textit{Cognitive Load:} High due to multitasking\\[3pt]

\textbf{Observable Signals:}\\
\textit{Facial Expression:} Determined\\
\textit{Body Posture:} Upright, attentive\\
\textit{Gaze Behavior:} Focused on the surroundings\\
\textit{Hand Motion:} Occasional gestures toward the dashboard\\
\textit{Behavior in Ride:} Occasionally monitoring the ride\\[3pt]

\textbf{Emergency Event:}\\
\textit{Type:} None\\
\textit{Description:} None\\
\textit{Severity:} None\\
\textit{Trigger Source:} None\\
\textit{Planning Impact:} None\\[3pt]

\textbf{LPS:}\\
\textit{Goal:} Catch a flight for a last-minute business trip\\
\textit{Destination:} Airport\\
\textit{Urgency:} High\\
\textit{Condition:} Stressed\\
\textit{Preference:} Efficiency\\
\textit{Risk Tolerance:} Medium\\[3pt]

\textbf{PIO:}\\
\textit{Trip Objective:} \textbf{Airport}\\
\textit{Driving Objective:} \textbf{Aggressive}\\
\end{minipage}
\\
\bottomrule
\end{tabular}
\caption{Qualitative example of PIR for a high-urgency airport trip prioritizing efficiency.}
\label{tab:pir_case_airport_aggressive}
\end{table*}

\begin{table*}[!t]
\centering
\small
\setlength{\tabcolsep}{6pt}
\renewcommand{\arraystretch}{1.1}
\begin{tabular}{@{}p{0.96\textwidth}@{}}
\toprule
\begin{minipage}[t]{\linewidth}
\textbf{Case 6: Nighttime School Pickup}\\[2pt]

\textbf{Explicit Intent:}\\
Talk: Oh no, my kid has been waiting alone at school for ages, and it's getting dark already! I really need to get there as fast as possible. Please hurry---what if they start feeling scared by themselves?\\[3pt]

\textbf{Implicit Intent:}\\
\textit{Scenario:} A mother is anxious about reaching her child, who is waiting alone at school as daylight fades. Traffic delays increase her stress and urgency.\\
\textit{Age Range:} 35--44\\
\textit{Gender:} Female\\
\textit{Occupation:} Manager\\
\textit{Passenger Frequency:} Almost never\\
\textit{Driving Experience:} More than 10 years\\
\textit{AV Experience:} No, but has used advanced driver-assistance systems\\
\textit{AV Trust Baseline:} Low\\
\textit{Passenger Monitoring Tendency:} Very high\\
\textit{Risk Attitude:} Somewhat cautious\\
\textit{Physical Condition Baseline:} Back pain or joint issues\\
\textit{Social Context:} Partner\\[3pt]

\textbf{Passenger State:}\\
\textit{Emotion:} Anxious, stressed, concerned\\
\textit{Physical State:} Back pain or joint issues\\
\textit{Comfort State:} Uncomfortable\\
\textit{Trust State:} Low trust in AV\\
\textit{Attention State:} High attention to the road\\
\textit{Cognitive Load:} High\\[3pt]

\textbf{Observable Signals:}\\
\textit{Facial Expression:} None\\
\textit{Body Posture:} Leaning forward; sitting rigidly or tensely\\
\textit{Gaze Behavior:} Looking out the window at traffic anxiously\\
\textit{Hand Motion:} Tapping fingers, fidgeting, or bouncing a leg\\
\textit{Behavior in Ride:} Repeatedly checking the phone or watch; sighing or taking deep breaths\\[3pt]

\textbf{Emergency Event:}\\
\textit{Type:} None\\
\textit{Description:} None\\
\textit{Severity:} None\\
\textit{Trigger Source:} None\\
\textit{Planning Impact:} None\\[3pt]

\textbf{LPS:}\\
\textit{Goal:} Pick up the child from school quickly\\
\textit{Destination:} School\\
\textit{Urgency:} High\\
\textit{Condition:} Anxious\\
\textit{Preference:} Efficiency\\
\textit{Risk Tolerance:} Medium\\[3pt]

\textbf{PIO:}\\
\textit{Trip Objective:} \textbf{School}\\
\textit{Driving Objective:} \textbf{Aggressive}\\
\end{minipage}
\\
\bottomrule
\end{tabular}
\caption{Qualitative example of PIR for an urgent school pickup prioritizing timely arrival.}
\label{tab:pir_case_school_aggressive}
\end{table*}

\clearpage
\onecolumn
\twocolumn

\subsection{PIO-Conditioned Planning}
\label{app:pio_conditioned_planning}

This subsection provides a controlled visualization of how PIO conditions downstream planning rather than evaluating a new planning architecture. PIO is supplied as structured passenger-intent context to the route and trajectory stages, allowing the planner to select and optimize feasible motions according to passenger needs while retaining the same environmental and safety constraints. To isolate this conditioning effect, the example below fixes both the driving scene and the trip objective and varies only the driving objective among \texttt{Aggressive}, \texttt{Standard}, and \texttt{Conservative}.

\begin{figure}[H]
  \centering
  \includegraphics[width=0.98\columnwidth]{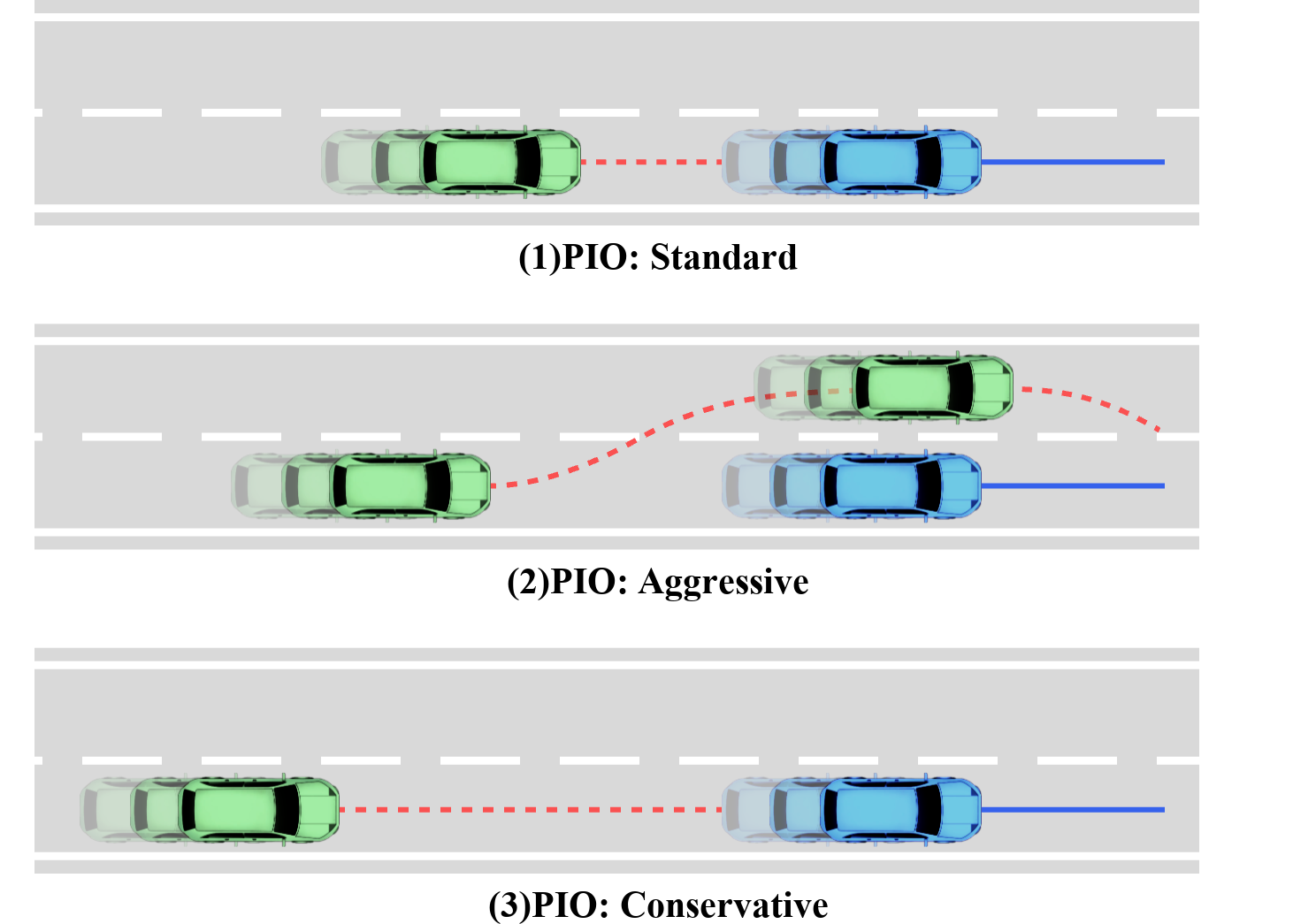}
  \caption{Illustrative demo of PIO-conditioned planning with the same driving scene and trip objective. Varying only the driving objective produces distinct behaviors: Aggressive favors faster progress and earlier overtaking, Standard balances progress and comfort, and Conservative preserves larger margins and smoother motion.}
  \label{fig:pio_planning_trajectories}
\end{figure}

Figure~\ref{fig:pio_planning_trajectories} shows that changing the PIO condition produces distinct but feasible behaviors under otherwise identical settings. The Aggressive condition initiates an earlier passing maneuver, Standard maintains balanced longitudinal progress, and Conservative retains a larger following gap with smoother motion. HICP serves only as a downstream application testbed in this demonstration: it consumes PIO during planning to verify that the structured intent representation can induce executable behavioral changes. The example therefore provides application-level evidence that PIO is an actionable planning condition, without positioning HICP itself as the primary methodological contribution.

\subsection{PIR Training and Reasoning Details}
\label{app:pir_training_details}

This section provides the implementation details of the PIR used in our experiments.

\paragraph{Input Prompt and Output Schema}

Each input contains an instruction and a \texttt{Passenger Intent} JSON object comprising explicit intent and implicit passenger-related information. Although the scenario is stored as a separate field during data construction and quality control, it is merged into the implicit portion of \texttt{Passenger Intent} before being provided to PIR. The instruction requires the model to return only valid JSON with exactly two top-level keys, \texttt{lps} and \texttt{pio}. The target follows the schema below:

{\small
\begin{verbatim}
{
  "lps": {
    "goal": {
      "purpose": "...",
      "destination": "..."
    },
    "urgency": "...",
    "condition": "...",
    "preference": "...",
    "risk_tolerance": "..."
  },
  "pio": {
    "trip_objective": "...",
    "driving_objective":
      "Aggressive|Standard|Conservative"
  }
}
\end{verbatim}
}

\paragraph{Joint LPS and PIO Generation}

PIR generates the LPS and PIO jointly in a single autoregressive response rather than through two separately trained models. The output is ordered such that the complete \texttt{lps} object precedes the \texttt{pio} object. Consequently, although training and inference are performed in one stage, generation of the PIO tokens is autoregressively conditioned on the preceding LPS tokens. This implementation preserves the conceptual LPS-to-PIO hierarchy while avoiding error propagation between independently decoded stages.

\paragraph{Training Objective}

We train PIR through SFT with LoRA. The objective is the standard token-level autoregressive cross-entropy over assistant response tokens:

\begin{equation}
\mathcal{L}_{\mathrm{SFT}}
=
-\sum_{t\in\mathcal{Y}}
\log p_{\theta}(y_t\mid x,y_{<t}),
\end{equation}

where $x$ is the input prompt and $\mathcal{Y}$ is the set of target response positions. Prompt tokens are masked with $-100$ and therefore do not contribute to the loss. The selected final checkpoint reached a training loss of $0.0463$ after 30 epochs.

All reported PIR results are averaged over three independent training runs with different random seeds. The three runs use the same train/validation/test splits, hyperparameters, and evaluation protocol; only stochastic initialization and data ordering differ. Tables in the main paper report the mean accuracy across these runs.

\paragraph{Training Hyperparameters}

\begin{table}[!htbp]
\centering
\small
\setlength{\tabcolsep}{4pt}
\begin{tabular}{ll}
\toprule
Hyperparameter & Value \\
\midrule
Backbone & Qwen3-4B \\
Fine-tuning method & LoRA-SFT \\
LoRA rank & 8 \\
LoRA $\alpha$ & 16 \\
LoRA dropout & 0.0 \\
Learning rate & $1\times10^{-4}$ \\
Training epochs & 30 \\
Per-device batch size & 1 \\
Gradient accumulation steps & 8 \\
Effective global batch size & 8 \\
Learning-rate scheduler & Cosine \\
Warmup ratio & 0.1 \\
\bottomrule
\end{tabular}
\caption{Training hyperparameters for PIR. The effective batch size is computed with one device.}
\label{tab:pir_training_hyperparameters}
\end{table}

\paragraph{Compute Resources and Budget}

For reproducibility and compute accounting, Table~\ref{tab:pir_compute_budget} records the hardware used for PIR training, the duration of a final run, and the total computational budget across all development and final runs. The GPU-hour estimate covers retained development and final jobs rather than only the final reported checkpoint.

\begin{table}[!htbp]
\centering
\small
\setlength{\tabcolsep}{4pt}
\begin{tabular}{p{0.34\columnwidth}p{0.56\columnwidth}}
\toprule
Compute item & Value \\
\midrule
GPU model and memory & NVIDIA H200 NVL (141 GB HBM3e) \\
Number of GPUs & 1 \\
Time per final training run & $\approx$1.6--2.2 h (30 epochs; 4B: $\approx$2.15 h) \\
Total GPU hours & $\approx$71.2 GPU-h \\
Budget coverage & Development + final runs \\
\bottomrule
\end{tabular}
\caption{Compute infrastructure and budget for PIR. GPU-hours include all retained development and final Slurm GPU jobs, including failed and cancelled attempts.}
\label{tab:pir_compute_budget}
\end{table}

\paragraph{Training and Evaluation Splits}

PIR is evaluated on the same held-out set of 240 real participant-derived samples used throughout the passenger-intent experiments; no synthetic samples are included in this test set. After holding out the test set, the remaining 2,000 samples are split 90/10 into 1,800 training and 200 validation samples. A manual cross-split audit confirms that no exactly duplicated record appears across the test and training/validation splits. The real/synthetic composition of each split is summarized in Table~\ref{tab:hpid_splits}. Hyperparameter selection and model development use only the training and validation portions, leaving the test set untouched until final evaluation.

\paragraph{Inference Configuration}

At test time, we use \texttt{max\_new\_tokens}$=768$, temperature $=0.2$, \texttt{top\_p}$=0.9$, and \texttt{top\_k}$=20$. The repetition penalty is left at the script default of $1.0$. Generation terminates using the Qwen chat-template end token, \texttt{<|im\_end|>}.

\paragraph{Output-Space Constraints}

We constrain the output space through the prompt and data normalization rather than grammar-constrained decoding. First, the instruction explicitly restricts \texttt{driving\_objective} to \texttt{Aggressive}, \texttt{Standard}, or \texttt{Conservative}. Second, labels expressed as \texttt{normal} during data construction are normalized to \texttt{Standard}. We do not apply a JSON-schema or grammar-level decoding constraint. Nevertheless, every generated \texttt{driving\_objective} on the test set belongs to one of the three valid categories.

\paragraph{Structured Prediction versus Free-Text Reasoning}

The PIR evaluated in this paper performs structured prediction: its response contains only the \texttt{lps} and \texttt{pio} JSON objects and does not include a free-text chain-of-thought field. We also explored a separate \texttt{think\_lps\_pio} task that combines free-text reasoning fields (e.g., \texttt{lps\_think} and \texttt{pio\_think}) with structured outputs, but that variant is not used for the results reported in this paper. Thus, the evaluated PIR is a one-stage LoRA-SFT model for joint structured LPS and PIO generation rather than a system whose predictions depend on an exposed free-text reasoning trace.

\subsection{HPID Data Collection and Participant Demographics}
\label{app:hpid_collection}

\paragraph{Recruitment and Consent}

We recruited adults aged 18 years or older through an open online call. Before participating, each participant signed an informed-consent form covering both the questionnaire and interview. The raw survey export contained 230 submissions from 226 unique participants. For the four participants with duplicate submissions, we retained the most recent response, yielding a final participant cohort of 226. All retained responses passed the embedded attention check. Data were collected between April 14 and June 11, 2026.

\begin{figure}[!htbp]
  \centering
  \includegraphics[width=0.93\columnwidth]{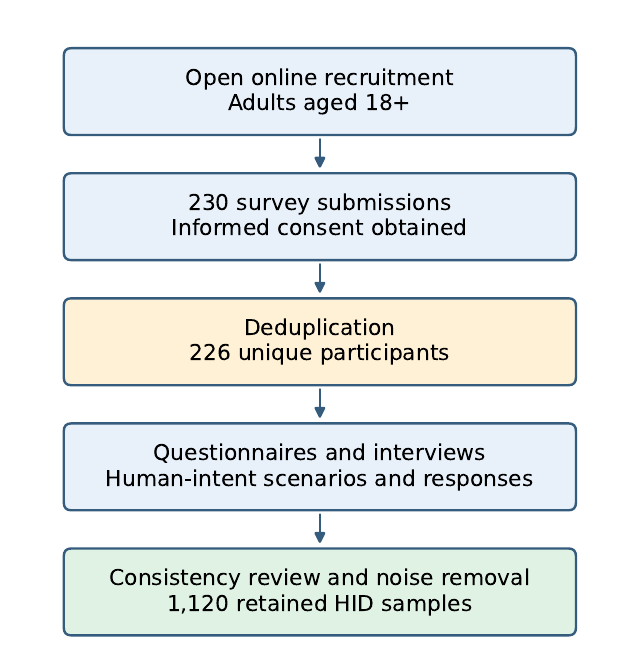}
  \caption{HPID data-collection and quality-control pipeline. Participant counts and dataset sample counts denote different units: multiple passenger-intent samples may be derived from one participant's questionnaire and interview.}
  \label{fig:hpid_collection_flow}
\end{figure}

\paragraph{Participant Demographics}

Table~\ref{tab:hpid_demographics} summarizes the final cohort. Participants covered all reported adult age groups and a broad range of occupations, including business and finance, management and administration, creative and design work, education, engineering and information technology, healthcare, service work, and full-time study. Most participants held a driver's license (87.6\%), while prior exposure to AVs varied substantially: 13.7\% had ridden in an AV, 26.5\% had used only advanced driver-assistance systems (ADAS), and 59.7\% reported neither experience.

\begin{table}[!htbp]
\centering
\small
\setlength{\tabcolsep}{4pt}
\begin{tabular}{lrr}
\toprule
Participant characteristic & $n$ & \% \\
\midrule
\multicolumn{3}{l}{\textit{Age}} \\
18--24 & 42 & 18.6 \\
25--34 & 82 & 36.3 \\
35--44 & 49 & 21.7 \\
45--54 & 38 & 16.8 \\
55--64 & 12 & 5.3 \\
65 or older & 3 & 1.3 \\
\midrule
\multicolumn{3}{l}{\textit{Gender}} \\
Man & 134 & 59.3 \\
Woman & 91 & 40.3 \\
Non-binary & 1 & 0.4 \\
\midrule
\multicolumn{3}{l}{\textit{Driving experience}} \\
No driver's license & 28 & 12.4 \\
Less than 1 year & 7 & 3.1 \\
1--5 years & 24 & 10.6 \\
5--10 years & 32 & 14.2 \\
More than 10 years & 135 & 59.7 \\
\midrule
\multicolumn{3}{l}{\textit{AV experience}} \\
Ridden multiple times & 22 & 9.7 \\
Ridden once & 9 & 4.0 \\
Used ADAS only & 60 & 26.5 \\
No prior experience & 135 & 59.7 \\
\bottomrule
\end{tabular}
\caption{Demographics and driving-related experience of the 226 unique participants. Percentages may not sum to 100 because of rounding.}
\label{tab:hpid_demographics}
\end{table}

\paragraph{Quality Control and Retained Samples}

The collected questionnaire and interview records were subjected to consistency review and noise removal. This process retained 1,120 passenger-intent samples for HPID. Figure~\ref{fig:hpid_collection_flow} distinguishes the number of participants from the number of retained samples; the latter is larger because each participant could contribute multiple intent scenarios. The annotation and quality-control procedure is detailed below.

\paragraph{Dataset Splits}

Table~\ref{tab:hpid_splits} reports the numbers of real and synthetic samples assigned to each dataset split. The held-out test set contains only real participant-derived samples; synthetic counterfactual samples are excluded from testing.

\begin{table}[!htbp]
\centering
\small
\setlength{\tabcolsep}{8pt}
\begin{tabular}{lrrr}
\toprule
Split & Real & Synthetic & Total \\
\midrule
Train & 792 & 1,008 & 1,800 \\
Validation & 88 & 112 & 200 \\
Test & 240 & 0 & 240 \\
\bottomrule
\end{tabular}
\caption{Numbers of real and synthetic samples in the HPID data splits. After reserving 240 real samples for testing, the remaining 2,000 samples are divided into training and validation sets using a stratified 90/10 split.}
\label{tab:hpid_splits}
\end{table}

\subsection{HPID Annotation Protocol and Quality Control}
\label{app:hpid_annotation}

\paragraph{Data Representation and Paired Construction}

Let the quality-controlled real-world dataset be
\begin{equation}
\mathcal{D}_{\mathrm{raw}}
=\{(s_i,h_i,c_i)\}_{i=1}^{N},
\end{equation}
where $s_i$ is a structured driving scenario, $h_i$ is the corresponding passenger-intent JSON object, and $c_i$ is the participant's self-reported driving-style choice from
$\mathcal{C}=\{\texttt{Aggressive},\texttt{Standard},\texttt{Conservative}\}$.

We construct a paired counterfactual dataset under a strict one-to-one rule: every real record produces exactly one generated counterpart. The counterpart retains the same scenario context, changes the driving-style label to a different class, and minimally revises only one to three leaf fields in the passenger-intent object. Therefore,
\begin{equation}
|\mathcal{D}_{\mathrm{aug}}|=2N.
\end{equation}
This minimal-edit policy preserves semantic locality between each real--counterfactual pair while introducing controlled, label-conditioned variation. With $N=1{,}120$ quality-controlled real samples, the resulting HPID contains 2,240 samples.

\paragraph{Stage A: One-to-One Counterfactual Augmentation}

For each real record, we sample a target class $c_i'\in\mathcal{C}\setminus\{c_i\}$ and prompt an LLM to revise only \texttt{passenger\_intent}. The generation must (1) preserve the JSON schema, (2) leave all scenario fields unchanged, (3) modify exactly one to three leaf attributes, and (4) include at least one modification from a priority evidence set, such as \texttt{explicit\_intent.talk}, emotional or physical state, posture or behavioral signals, or emergency-related fields. The counterfactual is assigned the paired identifier suffix \texttt{\_gen}, and the edited field paths are retained as metadata for subsequent validation.

\paragraph{Stage B: LPS Extraction}

GPT-4o extracts the structured LPS from the passenger's self-report $(s_i,h_i)$:
\begin{equation}
\begin{split}
\texttt{lps}=\{&\texttt{goal.purpose},\texttt{goal.destination},\\
&\texttt{urgency},\texttt{condition},\texttt{preference},\\
&\texttt{risk\_tolerance}\}.
\end{split}
\end{equation}
The output is normalized to the predefined schema and canonical value spaces; for example, urgency and risk tolerance are mapped to \texttt{low}, \texttt{medium}, or \texttt{high}.

\paragraph{Stage C: PIO Review}

A candidate PIO is initialized from the structured record: the trip objective follows the LPS destination, while the driving objective follows the participant-selected class $c_i$ for a real record or the target class $c_i'$ for its counterfactual counterpart. Passenger reviewers then inspect the complete trip experience, including the scenario, Passenger Intent, and LPS, and confirm or correct both objectives. The reviewed PIO serves as ground truth:
\begin{equation}
\begin{split}
\texttt{pio}=\{&\texttt{trip\_objective},\\
&\texttt{driving\_objective}\}.
\end{split}
\end{equation}
Here, \texttt{trip\_objective} specifies the destination-level objective, whereas \texttt{driving\_objective} is canonicalized to the three-class driving-style space. A counterfactual record is retained only when the reviewed driving objective agrees with its target class; otherwise, it is regenerated or removed.

\paragraph{Stage D: Reverse Reasoning Traces}

Finally, we condition on the scenario with its style label removed, the passenger-intent object, and the completed LPS and PIO annotations to generate concise explanation fields, \texttt{lps\_think} and \texttt{pio\_think}. These fields are reverse rationales that explain already specified structured targets; they are not used to predict or revise the target labels. The PIR results reported in the main paper use only the structured \texttt{lps} and \texttt{pio} targets, as described in Appendix~\ref{app:pir_training_details}.

\paragraph{Prompt Templates}

We use four task-specific protocols:
\begin{itemize}
  \item \textbf{Augmentation:} revise only \texttt{passenger\_intent} to match the target class, change one to three leaf fields, preserve the schema and scenario, and return JSON only.
  \item \textbf{LPS extraction:} use GPT-4o to extract an \texttt{lps} JSON object from the scenario and passenger self-report.
  \item \textbf{PIO review:} initialize the two objectives from LPS and the selected or target driving class, then let passenger reviewers confirm or correct them using the complete trip record; retain the reviewed PIO as ground truth.
  \item \textbf{Rationale generation:} given the scenario, passenger intent, and target LPS and PIO, return concise reverse reasoning traces in JSON format.
\end{itemize}
All model-generated records use JSON-only responses. Deterministic post-processing then normalizes field names, value spaces, and driving-style labels.

\paragraph{Consistency and Quality Control}

Each generated record is checked against its paired real record. Validation rejects records that alter the scenario, violate the schema, retain the original style label, edit fewer than one or more than three passenger-intent leaf fields, or fail to modify priority evidence. LPS outputs and reviewed PIO labels are additionally checked for missing keys, invalid categorical values, and, for counterfactual records, disagreement between the reviewed driving objective and target class. Records failing validation are regenerated or removed. The remaining records undergo manual consistency inspection for semantic coherence across the scenario, passenger-intent evidence, LPS, and PIO. This procedure controls structural validity and pairwise semantic consistency; we do not report an inter-annotator agreement coefficient because the protocol does not use independent redundant labels as its primary annotation mechanism.

\begin{algorithm}[t]
\caption{One-to-One Counterfactual Construction and Structured Annotation}
\label{alg:hpid_annotation_pipeline}
\begin{algorithmic}[1]
\footnotesize
\REQUIRE Raw records $\mathcal{D}_{\mathrm{raw}}=\{(s_i,h_i,c_i)\}_{i=1}^{N}$ and class set $\mathcal{C}$
\ENSURE Paired dataset $\mathcal{D}_{\mathrm{final}}$ with LPS, PIO, and rationale fields
\STATE $\mathcal{D}_{\mathrm{aug}}\leftarrow\emptyset$
\FOR{$i=1$ to $N$}
  \STATE Sample $c_i'$ uniformly from $\mathcal{C}\setminus\{c_i\}$
  \STATE $\hat{h}_i\leftarrow\textsc{GPT4o-Augment}(s_i,h_i,c_i')$
  \STATE $\hat{h}_i\leftarrow\textsc{ValidateAndNormalize}(h_i,\hat{h}_i)$
  \STATE $r_i^{\mathrm{orig}}\leftarrow(s_i,h_i,c_i)$
  \STATE $r_i^{\mathrm{gen}}\leftarrow(s_i,\hat{h}_i,c_i')$ with suffix \texttt{\_gen}
  \STATE $\mathcal{D}_{\mathrm{aug}}\leftarrow\mathcal{D}_{\mathrm{aug}}\cup\{r_i^{\mathrm{orig}},r_i^{\mathrm{gen}}\}$
\ENDFOR
\FORALL{$r\in\mathcal{D}_{\mathrm{aug}}$}
  \STATE $r.\texttt{lps}\leftarrow\textsc{GPT4o-LPS}(r.s,r.h)$
  \STATE $r.\texttt{lps}\leftarrow\textsc{CanonicalizeLPS}(r.\texttt{lps})$
  \STATE $r.\texttt{pio}\leftarrow\textsc{PassengerReviewPIO}(r)$
  \STATE $r.\texttt{pio}\leftarrow\textsc{CanonicalizePIO}(r.\texttt{pio})$
  \STATE $r.\texttt{traces}\leftarrow\textsc{LLM-Think}(r.s^{-c},r.h,r.\texttt{lps},r.\texttt{pio})$
\ENDFOR
\STATE \textbf{return} $\mathcal{D}_{\mathrm{aug}}$ as $\mathcal{D}_{\mathrm{final}}$
\end{algorithmic}
\end{algorithm}

\subsection{Additional PIO Baselines}
\label{app:additional_pio_baselines}

\paragraph{Comparison with Larger Models}

To strengthen the comparison beyond the pretrained Qwen3-4B baseline in the main paper, we evaluate three additional large models on exactly the same 240 held-out HPID samples. All models receive the same scenario and passenger-intent fields and are evaluated against the same reference annotations. Table~\ref{tab:additional_large_model_pio} reports their field-level accuracy together with the PIR results from Tables~\ref{tab:lps} and~\ref{tab:pio}; no joint-field score is included. As in the main evaluation, categorical fields use normalized exact match, while open-ended fields use GPT-4o semantic-equivalence judgments.

\begin{table}[!htbp]
\centering
\scriptsize
\setlength{\tabcolsep}{1.5pt}
\begin{tabular}{lcccc}
\toprule
Field & Gemma 4 & Kimi K2.5 & Qwen 3.7 & PIR (Ours) \\
\midrule
\texttt{lps.goal} & 69.17\% & \textbf{76.25\%} & 72.29\% & 69.79\% \\
\texttt{lps.urgency} & 61.25\% & 4.17\% & 0.83\% & \textbf{85.00\%} \\
\texttt{lps.condition} & 48.33\% & 63.75\% & 56.25\% & \textbf{76.67\%} \\
\texttt{lps.preference} & 59.58\% & 76.25\% & 68.33\% & \textbf{84.58\%} \\
\texttt{lps.risk\_tolerance} & 23.75\% & 2.08\% & 0.00\% & \textbf{90.42\%} \\
\texttt{pio.trip\_objective} & 67.08\% & 64.17\% & 50.83\% & \textbf{81.25\%} \\
\texttt{pio.driving\_objective} & 65.00\% & 57.92\% & 69.58\% & \textbf{85.42\%} \\
\bottomrule
\end{tabular}
\caption{Field-level LPS and PIO accuracy on the same 240-sample test set. PIR results are taken from Tables~\ref{tab:lps} and~\ref{tab:pio}. Goal accuracy for the additional baselines is the macro-average of purpose and destination accuracy, matching the aggregate reporting used for PIR. Bold denotes the best result in each row.}
\label{tab:additional_large_model_pio}
\end{table}

Table~\ref{tab:additional_large_model_pio} shows that Kimi K2.5 obtains the strongest aggregate goal accuracy, while the compact HPID-supervised PIR achieves the highest accuracy on the other six LPS and PIO fields. In particular, the low urgency and risk-tolerance scores of Kimi K2.5 and Qwen 3.7 indicate difficulty mapping free-form evidence into canonical categories. On trip and driving objectives, PIR exceeds the strongest additional baselines by 14.17 and 15.84 points, respectively. Thus, after HPID fine-tuning, the 4B model is highly competitive with substantially larger general-purpose models, reinforcing the value of task-specific supervision and explicit LPS-to-PIO modeling.

\clearpage
\paragraph{Effect of HPID across Qwen Model Scales}

To separate the effect of HPID supervision from that of model capacity, we further compare pretrained and HPID-fine-tuned variants within the Qwen3 family. For every model size, the two variants use the same backbone and are evaluated on the same 240 held-out HPID samples; they differ only in whether SFT is performed on HPID. Following Table~\ref{tab:additional_large_model_pio}, we report accuracy for each of the five LPS attributes and both PIO fields rather than an aggregate score. This paired, field-level design reveals both the overall benefit of HPID and which intent dimensions benefit most at each model scale.

\begin{table}[H]
\centering
\scriptsize
\setlength{\tabcolsep}{2pt}
\resizebox{\columnwidth}{!}{%
\begin{tabular}{lcccccc}
\toprule
& \multicolumn{2}{c}{Qwen3-0.6B} & \multicolumn{2}{c}{Qwen3-1.7B} & \multicolumn{2}{c}{Qwen3-4B} \\
\cmidrule(lr){2-3}\cmidrule(lr){4-5}\cmidrule(lr){6-7}
Field & Pretrained & +HPID & Pretrained & +HPID & Pretrained & +HPID \\
\midrule
\texttt{lps.goal} & 3.44 & 65.83 & 0.21 & 67.08 & 5.63 & 69.79 \\
\texttt{lps.urgency} & 38.75 & 80.42 & 35.83 & 83.75 & 60.42 & 85.00 \\
\texttt{lps.condition} & 13.33 & 70.42 & 12.50 & 74.58 & 17.50 & 76.67 \\
\texttt{lps.preference} & 7.92 & 83.75 & 18.33 & 80.83 & 52.92 & 84.58 \\
\texttt{lps.risk\_tolerance} & 8.33 & 88.33 & 4.58 & 85.83 & 9.17 & 90.42 \\
\texttt{pio.trip\_objective} & 10.42 & 72.50 & 36.25 & 74.17 & 37.50 & 81.25 \\
\texttt{pio.driving\_objective} & 11.25 & 82.92 & 40.83 & 82.08 & 60.83 & 85.42 \\
\bottomrule
\end{tabular}
}
\caption{Field-level effect of HPID supervision across Qwen3 model scales on the same 240-sample test set. Values are accuracy percentages; \texttt{lps.goal} is the macro-average of purpose and destination accuracy. Each pretrained model is paired with an HPID-supervised variant using the identical backbone.}
\label{tab:qwen_scaling_hpid}
\end{table}

\clearpage

\end{document}